\documentclass[apj]{emulateapj}

\usepackage{graphicx}
\usepackage{times}
\usepackage{amsmath}
\usepackage{color}
\usepackage{graphics}
\usepackage{subfigure}
\usepackage{txfonts}
\usepackage{natbib}

\bibpunct{(}{)}{;}{a}{}{,}

\shorttitle{Observational evidence of torus instability}
\shortauthors{F.P. Zuccarello et al.}

\begin{document}


         
\title{{Observational evidence of torus instability as trigger mechanism for coronal mass ejections:
         the 2011 August 4 filament eruption  }}

\author{F.P. Zuccarello\altaffilmark{1,2}}
\email{Francesco.Zuccarello@wis.kuleuven.be}
\author{D.B. Seaton\altaffilmark{2}}
\email{dseaton@oma.be}
\author{M. Mierla\altaffilmark{2,3}}
\email{marilena@oma.be}
\author{S. Poedts\altaffilmark{1}}
\email{Stefaan.Poedts@wis.kuleuven.be}
\author{L. A. Rachmeler\altaffilmark{2}}
\email{rachmeler@oma.be}
\author{P. Romano\altaffilmark{4}}
\email{Paolo.Romano@oact.inaf.it}
\author{F. Zuccarello\altaffilmark{5}}
\email{fzu@oact.inaf.it}

\altaffiltext{1}{Centre for mathematical Plasma-Astrophysics, KU Leuven, Celestijnenlaan 200B, 3001 Leuven, Belgium.}
\altaffiltext{2}{Royal Observatory of Belgium, Ringlaan 3, 1180 Brussels, Belgium}
\altaffiltext{3}{Institute of Geodynamics of the Romanian Academy Jean-Louis Calderon 19-21, RO-020032,Bucharest, Romania}
\altaffiltext{4}{INAF - Osservatorio Astrofisico di Catania, via S. Sofia 78,  95123 Catania, Italy}
\altaffiltext{5}{Dipartimento di Fisica e Astronomia - Universit\'{a} di Catania Via S.Sofia 78, 95123 Catania, Italy}

\begin{abstract}
Solar filaments are magnetic structures often observed in the solar atmosphere and consist of plasma that is cooler and denser than their surroundings.
They are visible for days -- and even weeks -- which suggests that they are often in equilibrium with their environment before disappearing or erupting.
Several eruption models have been proposed that aim to reveal what mechanism causes (or triggers) these solar eruptions. Validating these models through observations represents a fundamental step in our understanding of solar eruptions. 
We present an analysis of the observation of a filament eruption that agrees with the torus instability model. This model predicts that a magnetic flux rope embedded in an ambient field undergoes an eruption when the axis of the flux rope reaches a critical height that depends on the topology of the ambient field.  
We use the two vantage points of SDO and STEREO to reconstruct the three-dimensional shape of the filament, to follow its morphological evolution and to determine its height just before  eruption. The magnetograms acquired by SDO/HMI are used to infer the topology of the ambient field and to derive the critical height for the onset of the torus instability. 
Our analysis shows that the torus instability is the trigger of the eruption. We also find that some pre-eruptive processes, such as magnetic reconnection during the observed flares and flux cancellation at the neutral line, facilitated the eruption by bringing the filament to a region where the magnetic field was more vulnerable to the torus instability. 

\end{abstract}

\keywords{Sun: coronal mass ejections (CMEs)  --- Sun: corona --- Sun: filaments, prominences --- methods: observational}

\section{Introduction}

\begin{table*} 
\caption{List of the goes flares. The key events in the  evolution of the active region are in boldface.}             
\label{table1}
\centering    
\renewcommand{\arraystretch}{1.5}
\begin{tabular}{ c  c p{0.2cm} c p{0.5cm} l}    
\hline\hline                 
      \textbf{Day}    &  \textbf{Time (UT) } & &    \textbf{ GOES Flare (Beg, Max, End)}    &   & \multicolumn{1}{c}{\textbf{Description}} \\ 
\hline                                                                                     
  \textbf{2011-08-03} &   \textbf{03:40}     & &    \textbf{M1.1 (03:08,  03:37,  03:51)}   &   &  \textbf{Figure 1A: the two filaments are visible.}  \\
          2011-08-03  &           04:29      & &            M1.7 (04:29,  04:32,  04:35)    &   &  Flare occurred in the nearby active region. \\ 
  \textbf{2011-08-03} &   \textbf{06:42 }    & &    \textbf{C1.1 (06:42,  06:46,  06:49)}   &   &  \textbf{Figure 1B: the small filament is activated.}  \\
          2011-08-03  &           07:38      & &            C8.7 (07:38,  07:58,  08:06)    &   &  Occurred in the southern part of the active region. \\
          2011-08-03  &           10:01      & &            C1.0 (10:01,  10:04,  10:06)    &   &  Occurred in the southern part of the active region. \\
  \textbf{2011-08-03} &   \textbf{13:30 }    & &   \textbf{M6.0 (13:17,  13:48,  14:10)}   &   &  \textbf{Figure 1C: part of the small filament erupts.}    \\
  \textbf{2011-08-03} &   \textbf{15:08 }    & &          \textbf{ ---}                    &   &  \textbf{Figure 1D: morphological change of the bigger       filament. }   \\
          2011-08-03 &            18:52      & &  C2.3 (18:52,  19:00,  19:06)            &   &  Flare occurred in the nearby active region. \\ 
 \textbf{2011-08-03} &   \textbf{20:02}      & &  \textbf{C8.5 (19:23,  19:30,  19:42)}   &   &  \textbf{Figure 1E: flare occurred close to the filament  foot point.} \\
          2011-08-04 &            00:47      & &  C1.7 (00:47,  01:09,  01:23)   &   &  Occurred in the southern part of the active region. \\
          2011-08-04 &            02:14      & &  C3.2 (02:14,  02:23,  02:28)   &   &  Occurred in the southern part of the active region. \\
  \textbf{2011-08-04} &   \textbf{03:48}     & &   \textbf{M9.3 (03:41   03:57   04:04) }  &   &  \textbf{Figure 1F: the bigger filament erupts.}  \\
 
\hline        
\hline                       
\end{tabular}

\end{table*}


Solar filaments are dark, filamentary structures constituted of plasma that is cooler and denser than their surroundings. Although the precise topology of the magnetic configuration suitable for supporting solar filaments is still under debate, magnetic flux ropes, with their shear, twist and the presence of dips, are among the most promising candidates \citep{Mac2009}.

Solar filaments are generally visible for days --- or even weeks --- before disappearing or erupting, which suggests that most of the time they are in equilibrium with their environment. The eruption of a filament is the manifestation of the sudden loss of this equilibrium. Understanding what causes (or triggers) the disruption of the equilibrium and drives filament eruptions and their associated coronal mass ejections (CMEs) is still an area of active research.

Several CME initiation models have been proposed \citep[see ][for a review]{Forbes2000,Klim2001, Rou2006, Forbes2006,Forbes2010,Chen2011}. Among the different CME initiation scenarios, one that relys on an ideal-MHD instability to trigger the eruption is the torus instability model \citep{Bat1978,Kliem2006}. In this model, a current ring of major radius $R$ is embedded in an external magnetic field. Current rings are subjected to a radially, outward directed hoop force that decreases when the ring expands. If the inward directed Lorentz force due to the external field decreases faster with $R$ than the hoop force, the system is unstable to perturbations. 


Assuming an external field $B_{ex} \propto R^{-n}$, where $n=-R\; d\left( \ln B_{ex} \right)$ $/dR$, \cite{Bat1978} showed such an instability will occur when $n>n_{crit} =1.5$. In other words, if the current ring has a major radius R such that the decay index of the external field, $n$, is significantly smaller than $n_{crit}$, the system is in a stable equilibrium where the inward magnetic tension of the external field balances the outward magnetic pressure of the current channel. However, when $n$ approaches $n_{crit}$ this equilibrium becomes unstable and any displacement of the current channel due to some perturbation will initiate an outward motion of the current ring.

\cite{Tor2007} performed numerical magnetohydrodynamic (MHD) simulations of Titov \& D\'{e}moulin-type flux ropes \citep[T\&D;][]{Tit1999} embedded in an external field. \citeauthor{Tor2007}  considered semicircular flux ropes with a small aspect ratio 
and confirmed that the torus instability occurs when the flux rope axis reaches a height where the decay index of the external field is larger than $n_{crit}$. They also found that the steeper the increase of the decay index --- that is, the faster the external field decreases with height --- the larger the initial acceleration of the flux rope is.  

Additionally, \cite{Dem2010} have shown that the torus instability model is the three-dimensional counterpart of the catastrophic loss of equilibrium model firstly introduced by \cite{van1978} and later generalized by \cite{For1991} and \cite{Ise1993} to include  the effect of finite coronal currents and photospheric line-tying. Starting from the model of \cite{Ise1993}, \cite{For1995} studied a two dimensional coronal flux rope embedded in the field generated by two photospheric magnetic sources, which undergo convergence motion towards the polarity inversion line that separates them. These authors found that as the distance between the two magnetic sources decreases, the magnetic energy stored in the flux rope increases until the system reaches a critical point at which it experiences a loss of equilibrium and erupts. When the loss of equilibrium occurs, a current sheet is formed below the flux rope that eventually finds a new equilibrium. If magnetic reconnection is allowed, the flux rope undergoes a full eruption \citep{Lin2000}. 

Flux cancellation at the polarity inversion line (PIL) due to convergence motions has also been studied using three-dimensional MHD simulations \citep{Ama2003,Ama2011,Zuc2012c}. In these models, magnetic reconnection occurs between highly sheared field lines in the proximity of the PIL as a consequence of the convergence flows. This reconnection allows the system to change from an arcade-like to a flux rope-like configuration that eventually experiences a full eruption. As the eruption progresses, the  magnetic reconnection below the flux rope, generally referred to as tether-cutting reconnection, has two effects: it reduces the magnetic tension of the overlying field and it also increase the magnitude of the poloidal field of the flux rope.

\cite{Aul2010} performed a similar simulation, but the effect of the flux dispersion was obtained by means of increased photospheric diffusion. The authors of that study found that tether-cutting reconnection is important for forming the flux rope, and facilitates its slow rise, but the actual trigger of the eruption is the torus instability. In fact, if the tether-cutting reconnection is stopped before the critical height for the onset of the torus instability is reached, no eruption is observed. Once the flux rope begins to accelerate, magnetic reconnection begins to occur below it, eventually transferring overlying field into flux rope field and resulting in a positive feedback that leads to a full eruption.  

Tether-cutting reconnection is not the only mechanism that can facilitate solar eruptions, however. For example, \cite{Sea2011} observed a flow of cold plasma in the low corona just prior to the eruption that occurred on 2010~April~3.  Using stereoscopic triangulation on observations from SWAP onboard PROBA2 and SECCHI onboard STEREO \citeauthor{Sea2011} reconstructed the three-dimensional evolution of the event and concluded that the initial mass off-loading process facilitated the rise of the flux rope, but the eruption itself was likely triggered by the catastrophic loss of equilibrium of the flux rope. \cite{Zuc2012b} extended this analysis by investigating the evolution of the magnetic field. \citeauthor{Zuc2012b} concluded that the eruption was compatible with the torus instability scenario --- that is, they found that the estimated location of the flux rope axis was in a region where the decay index was close to $n_{crit}$. As a result, the increase in the height of the flux rope after the mass off-loading may have been critical to facilitating the flux rope's crossing of the instability threshold, eventually resulting in the full eruption of the filament. 

\begin{figure*}
\begin{center}
\includegraphics[width=.95\textwidth]{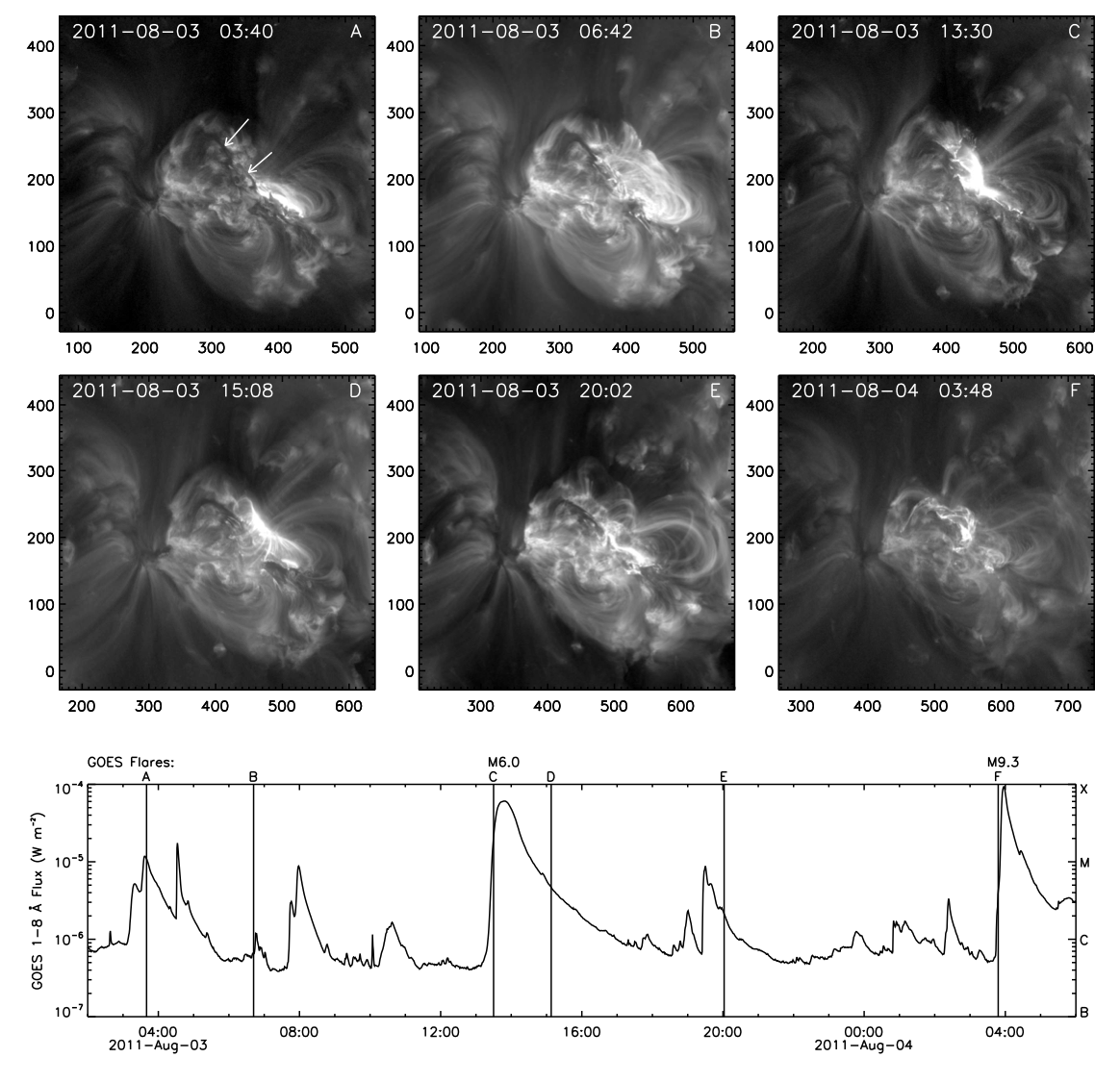}
\caption{Sequence of co-aligned SDO/AIA 193~\AA\, images acquired between 2011~August~3 03:40~UT and 2011~August~4 03:48~UT (top rows) and integrated X-ray (1-8~\AA) flux as measured by the X-ray Flux Monitor onboard the GOES~15 spacecraft between 2011~August~3 00:00~UT and 2011~August~4  06:00~UT (bottom panel). The vertical lines in the GOES flux indicate the times at which the AIA 193~\AA\, images were obtained.  The white arrows in the top left image indicate the locations of the two filaments. North is at the top of the image and west to the right. }
\label{event}
\end{center}
\end{figure*}

\begin{figure*}
\begin{center}
\includegraphics[width=.95\textwidth]{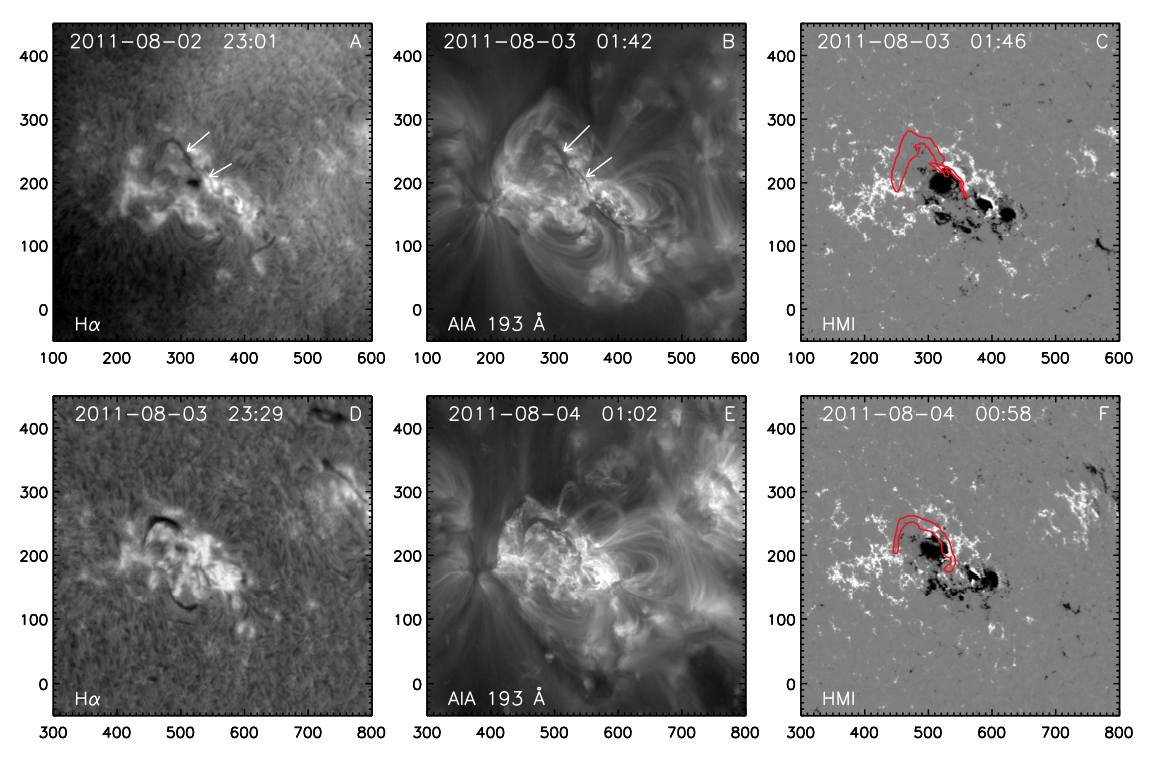}
\caption{Co-aligned images of the active region taken before (top panels) and after (bottom panels) the M6.0 flare that occurred on 2011~August~3 at 13:17~UT. Left panels: H$_{\alpha}$ images acquired by the BBSO. Middle panels: SDO/AIA images acquired at 193~\AA. Right panels: SDO/HMI magnetograms. The white arrows in the top images indicate the location of two filaments. The red contours on the HMI magnetograms outline the shape of the analyzed filaments derived from the AIA 193~\AA\, images. North is at the top of the images and west to the right. }
\label{morph}
\end{center}
\end{figure*}

In this paper, we present the analysis of a filament eruption that occurred on 2011~August~4 in NOAA active region (AR) 11261 whose interpretation supports the eruption scenario described by \cite{Aul2010} and \cite{Zuc2012c}. 

AR~11261 was composed of a group of three sunspots characterised by strong coherent negative polarities. The sunspots were trailed to the east by an extended facula associated with a positive magnetic flux distribution. Also, to the west of the northernmost sunspot there were several pores associated with positive magnetic field. The active region contained several filaments, but in this study we focus on the filament located along the PIL between the northernmost negative sunspot and the facula. 

On 2011~August~3 at 13:17~UT (hereafter referred to as August~3 at 13:17) an M6.0 flare occurred very close to the western footpoint of the filament without destabilizing it. Our analysis shows that at the moment of the M6.0 flare, the filament was torus stable. After the flare, we observed a change in the morphology of the filament as its western footpoint moved southward.

On August~4 at 03:48 the filament erupted completely. In the 14 hours between the M6.0 flare and the filament eruption, we observed flux cancellation at the PIL driven by converging flows. During this flux cancellation process we also observed several brightenings in the hotter SDO/AIA channels, suggesting the occurrence of tether-cutting reconnection driven by flux cancellation just like in the simulations of \cite{Aul2010} and \cite{Zuc2012c}.  

Combining three-dimensional reconstructions of the filament and potential magnetic field extrapolations we show in this paper that the trigger of the eruption was the torus instability. Our analysis also shows that the change in the morphology of the filament and the observed flux cancellation were fundamental to facilitating the eruption. In fact, due to this morphological change, the filament extended further south, into a region where the filament was more vulnerable to the torus instability. In this new configuration, the flux cancellation, as suggested by the simulation of \cite{Aul2010}, removed part of the line-tying allowing the rise of the filament up to the height where the decay index is larger then $n_{crit}$, eventually resulting in an eruption.

\section{Observations} \label{Sec:Obs}

In order to investigate the coronal evolution of active region AR~11261, we used images acquired by the Atmospheric Imaging Assembly \cite[AIA;][]{AIA} on board the Solar Dynamic Observatory \cite[SDO;][]{SDO} with a pixel resolution of about 0.6~arcsec and a cadence of 12~s. For this study, we used images acquired between August~3 at 01:42 and August~4 at 03:48 in the 304~\AA,\, 193~\AA\, and 131~\AA\, passbands.

The face-on view of the sun provided by SDO/AIA is complemented by views from the side --- with increasing angular separation over the past few years --- from the Extreme Ultraviolet Imagers \citep[EUVI;][]{Wue2004} on board the twin Solar Terrestrial Relations Observatory \cite[STEREO;][]{Kai2008} spacecraft. On the day of the eruption, the separation angles between Earth and STEREO-A and -B were 100$^{\circ}$ and 93$^{\circ}$, respectively. Since the erupting filament was located at a longitude of about 37$^{\circ}$~W, the eruption was entirely obscured by the limb of the sun from the vantage point of STEREO-B. Therefore, we only used the observations provided by STEREO-A. In particular, we used images acquired by EUVI-A between August~3 at 06:35 and August~4 at 03:46 in the 195~\AA\, and 304~\AA\ passbands. EUVI-A images have a pixel resolution of about 1.6~arcsec, and, on the date of the eruption, were acquired with a time cadence of 5~min in the 195~\AA\, passband and 10~min in the 304~\AA\, passband.

We also used full-disk H$_{\alpha}$ images with a spatial resolution of 1~arcsec acquired by the Big Bear Solar Observatory (BBSO) on August~2 and August~3 to infer the morphology of the filament. 

Finally, to analyze the magnetic configuration of the active region, we used  full-disk line-of-sight magnetograms from the Heliospheric and Magnetic Imager \cite[HMI;][]{HMI} at 6767.8~\AA\, with a pixel resolution of 0.5 arcsec and a temporal resolution of 12 min. We used magnetograms acquired between August~3  at 00:00 and August~4 at 03:48. All the magnetogram data were corrected for the angle between the magnetic field direction and the observer's line-of-sight and were co-aligned by applying the standard differential rotation rate reported by \cite{How1990}.

\section{Analysis}\label{Sec:Analysis}

AR~11261 first appeared on the east limb of the Sun on July~27 and, during its passage across the solar disk, was the source of several C- and M-class flares. Table~\ref{table1} includes a list of the flares that occurred during the period of observation, from August~3 at 02:00 until August~4 at 04:00. The key events in the evolution of the active region are indicated in boldface. Figure~\ref{event} shows the GOES X-ray flux \citep{Ste2004} for the same time interval together with representative AIA 193~\AA\, images. The acquisition times of the AIA images shown in the upper rows are indicated by vertical lines in the GOES flux plot, corresponding to the boldface rows in Table~1.  

Figure~\ref{event}A shows the configuration of the active region on August~3 at 03:40. At this time, two filaments (indicated by arrows) are visible: a big cusp-shaped filament and smaller one to the south-west of it.

On August~3 at 06:42 the smaller filament was activated (Fig.~\ref{event}B) and part of it eventually erupted, on August~3 at 13:17, resulting in a halo CME and in an M6.0 flare (Fig.~\ref{event}C) that is characterized by a long decay phase and by the presence of intense post flare loops (Fig.~\ref{event}D). 

As we will show in the following subsections, this M6.0 flare represents a milestone in the evolution of the active region. Following the M6.0 flare, there is a clear change in the bigger filament's morphology. The now-larger filament extended farther south into the region where the smaller filament was previously observed (Fig.~\ref{event}D--\ref{event}E). The filament remained in this new configuration until August~4 at 03:48 when it was activated and underwent a complete eruption. Figure~\ref{event}F shows the early phase of the eruption, highlighting the twisted structure of the flux rope that probably supported the filament material. 

\begin{figure}
\begin{center}
\includegraphics[width=.45\textwidth]{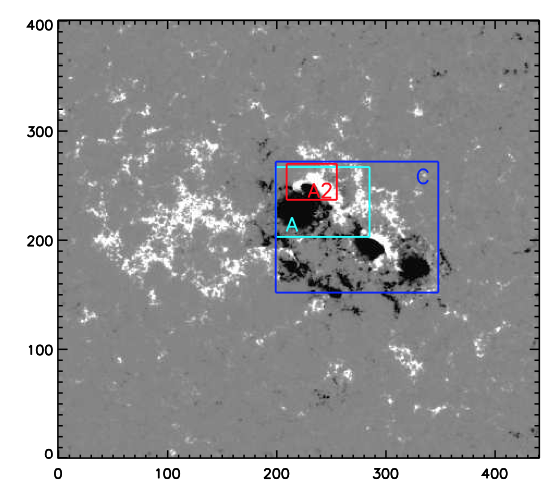}
\caption{HMI magnetogram of the active region obtained on 2011~August~3 at 00:00~UT. The three squares indicate the three different regions discussed in the text. Region C is the region that includes the complete sunspot group. Region A is a zoom into the northern sunspot, while Region A2 is a zoom into the locations where the footpoints of the two filaments cross each other. The $x$ and $y$ scales are in image pixels. North is at the top and west to the right.}
\label{flux-regions}
\end{center}
\end{figure}

\begin{figure*}
\begin{center}
\includegraphics[width=.95\textwidth]{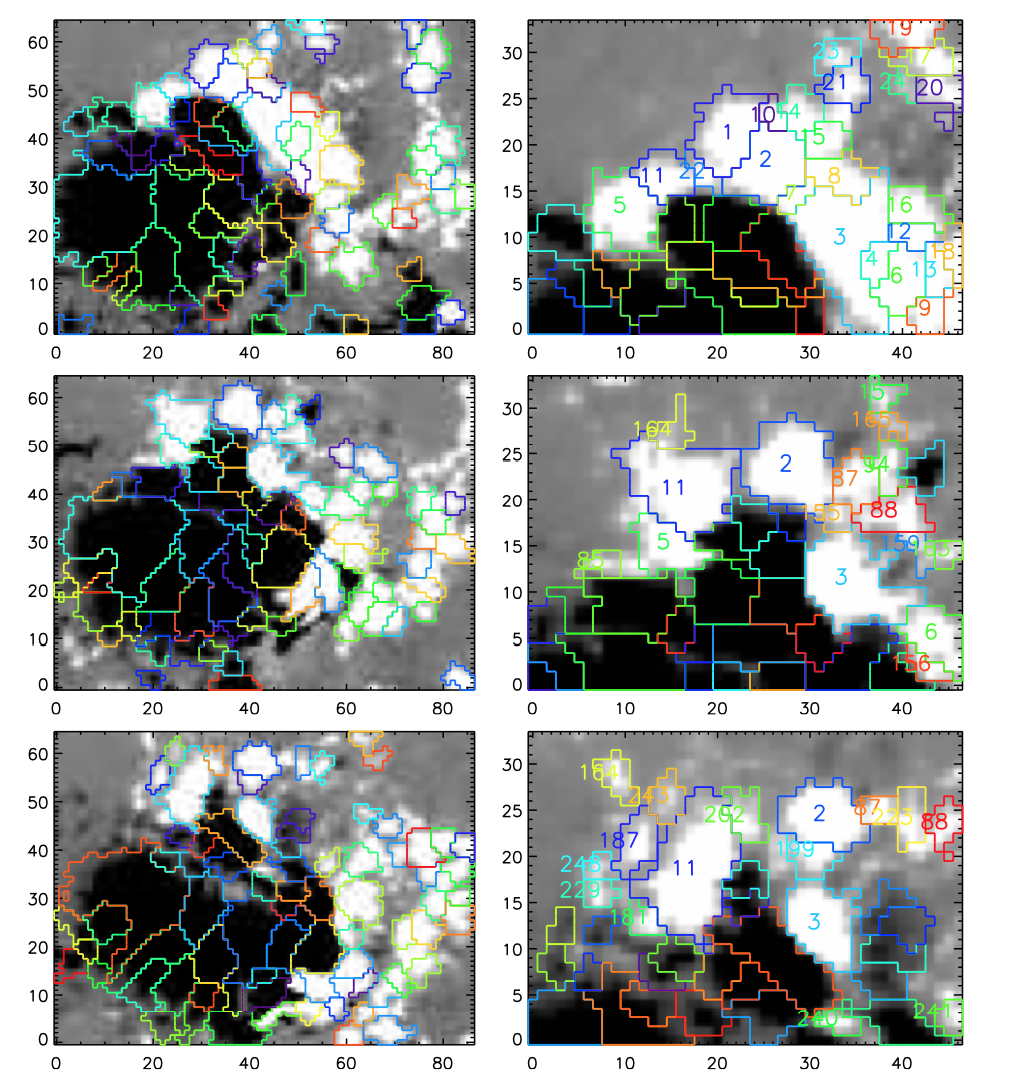}
\caption{Zoom of the HMI magnetograms in Region A (left panels) and Region A2 (right panels) at different times: 2011~August~3 at 00:00~UT (top panels), August~3 at 18:00 (middle panels) and August~4 at 03:24 (bottom panels). The color contours highlight the features identified using the YAFTA algorithm. For clarity, we show only a sample of the magnetic feature labels: those corresponding to the positive flux in Region A2. The magnetic scale is saturated at $\pm$ 300 G. The $x$ and $y$ scales are in image pixels. North is at the top and west to the right.}
\label{Yafta}
\end{center}
\end{figure*}

\begin{figure}
\begin{center}
\includegraphics[width=.45\textwidth]{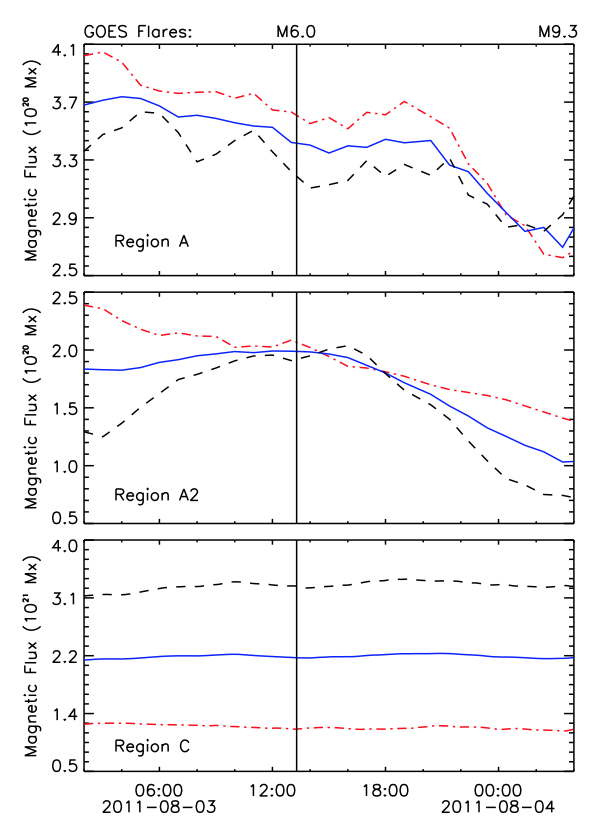}
\caption{Time evolution of the positive (dash-dotted red), negative (dashed black) and total unsigned (solid blue) magnetic flux for Region A (top), Region A2 (middle) and Region C (bottom).}
\label{flux}
\end{center}
\end{figure}

\subsection{Filament Morphology}

The most significant change in the morphology of the filament occurred during the M6.0 flare, so to further investigate the evolution of the filament, we considered images taken 12~h before and after the flare. Figure~\ref{morph} shows the H$_{\alpha}$ and the AIA 193~\AA\, images of the active region together with the relevant HMI magnetograms before (top panels) and after (bottom panels) the M6.0 flare. The images have been co-aligned using the mapping software available in \textsf{SolarSoft}. 

The H$_{\alpha}$ image acquired on August~2 at 23:01 shows the presence of a cusp-shaped filament that extends from the eastern facula up to the north of the sunspot. To the west of the sunspot, a smaller filament is visible as a dark threaded structure that crosses the facula. Our analysis of contemporaneous AIA images confirms the presence of two separate filaments. The western footpoint of the bigger filament seems to cross the eastern foot point of the smaller filament.

Figure~\ref{morph}C shows the contours of the filaments derived from the 193~\AA\, image (red lines) superimposed on the HMI magnetogram. The eastern footpoint of the bigger filament is anchored in the dispersed positive flux distribution of the eastern facula, while the western footpoint is anchored in the negative polarity of the sunspot. Meanwhile, the eastern footpoint of the smaller filament seems to be anchored in the positive flux distribution of the northern sunspot, while its western footpoint appears to be anchored in a negative flux intrusion at the south of it. 

The crossing of the footpoints of the two filaments, together with their magnetic configuration, may suggest that the two filaments are actually part of the same complex flux rope in which a significant amount of flux is anchored in the photosphere, close to the location where  the two filaments cross each other. 

The morphology of the active region after the M6.0 flare is shown in the bottom panels of Fig.~\ref{morph}. The H$_{\alpha}$ image is saturated by the emission of the facula, so it is not possible to clearly determine the morphology of the filament across the facular region. However, the 193~\AA\, image shows the presence of a  single arch-shaped filament that extends from the positive polarity of the eastern facula to the negative polarity at the south end of the northernmost sunspot. In this new configuration, the axis of the filament extends south along the PIL between the negative flux of the sunspot and the positive flux to the west of it, where the eastern leg of the smaller filament was anchored before the M6.0 flare.

\subsection{Magnetic flux}

During the 28 h that preceded the filament eruption, the photospheric magnetic field evolved considerably, especially close to the location where the footpoints of the two filaments cross each other (see Fig. \ref{morph}C--\ref{morph}F). In order to analyze the evolution of this magnetic field, in the sequence of co-aligned magnetograms we identified the three subregions that are shown in Fig. \ref{flux-regions}.

The region enclosing all the sunspots of the active region is labelled as Region C, while Region A is a subfield of Region C centred around the northern sunspot. Finally, the location where the footpoints of the filaments cross each other is labelled as Region A2. 

In Region C we calculated the positive, negative and total unsigned fluxes by integrating over the full region. However, this approach could not be used for Region A and Region A2. In fact, the magnetic field in these regions is quite dynamic, so during the 28~h during which we tracked the field evolution some of the flux left the region, passing through the lateral boundary. To overcome this difficulty, we used the YAFTA (Yet Another) Feature Tracking Algorithm \citep{Wel2003} to identify and follow the different magnetic features. In this algorithm, a magnetic structure is identified as a feature only if the magnetic field exceeds a threshold of 50 Gauss and extends to at least 16 pixels for Region A or 8 pixels for Region A2. Among all the identified features we consider only those that are tangent to the PIL at step zero --- that is, on August~3 at 00:00 --- and that do not leave the subfield until August~4 at 03:24. However, if one of the tracked features disappears due to either fragmentation or merging and a new feature appears in its place, this latter is also included in our analysis. We then computed the total, positive and negative flux by summing the magnetic field of all the features that meet the aforementioned requirements at each time step.

Figure~\ref{Yafta} shows the features found using YAFTA as well as the respective magnetogram for Regions A (left panels) and A2 (right panels). 
A significant amount of magnetic flux is cancelled at the PIL and especially along the northern part of the PIL. This is confirmed by Fig.~\ref{flux}, which shows the total (solid blue), positive (dash-dotted red) and negative (dashed black) magnetic flux as a function of time for Region A (top), Region A2 (middle) and Region C (bottom). The vertical line indicates the time of the M6.0 flare; the M9.3 flare occurs just at the end of the observation. 
 
In Region A the negative flux increases until August~3 at 05:00 and then decreases slowly until 20:00 on the same day when the decrease becomes much steeper. This trend continues until the onset of the eruption. The flux evolution follows a similar trend in Region A2, where the steep decrease in the positive and negative flux begins only a couple of hours after the M6.0 flare. On the other hand, the flux in Region C behaves completely differently. Both the positive and negative magnetic fluxes remain almost constant throughout the period of observation, though the negative flux increases very slightly. 

There is a significant magnetic flux imbalance evident in Figure~\ref{flux}: the negative flux amounts to roughly twice the positive flux. Note that this flux imbalance disappears, however, if we take the dispersed magnetic field in the eastern facula into consideration.

\subsection{Velocity field}

\begin{figure}
\begin{center}
\includegraphics[width=.45\textwidth]{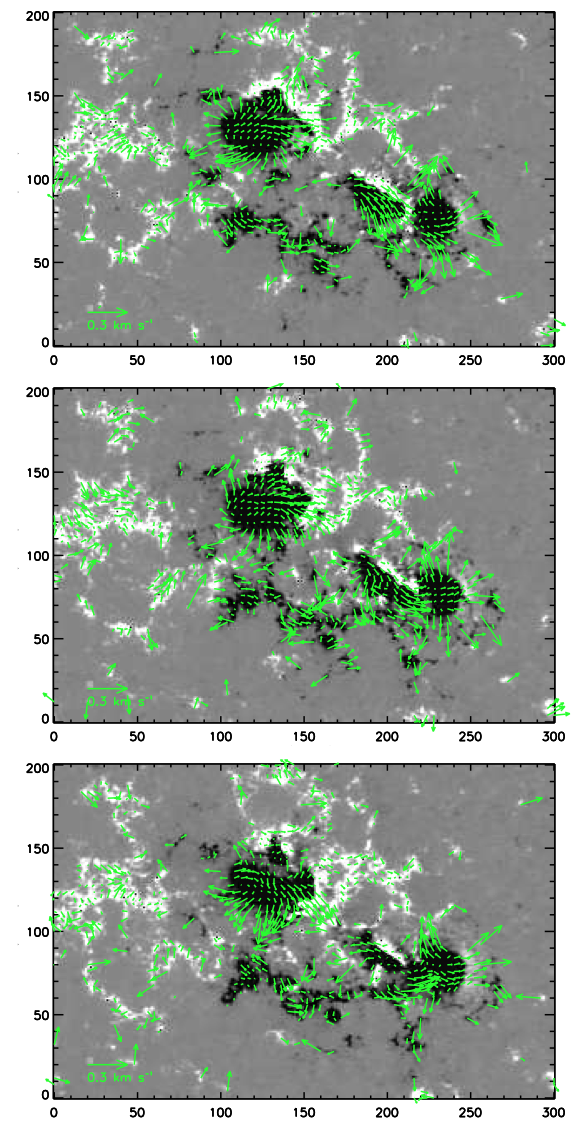}
\caption{Velocity field derived using the DAVE algorithm for 2011~August~3 11:12~UT (top),  2011~August~3 19:12~UT (middle) and  2011~August~4 02:36~UT (bottom). The $x$ and $y$ scales are in image pixels. North is at the top and west to the right. }
\label{vel}
\end{center}
\end{figure}

In order to derive the photospheric velocity flows, we processed the sequence of co-aligned HMI magnetograms taken between August~3 at 00:00 and August~4 at 03:48 with the Differential Affine Velocity Estimator \citep[DAVE;][]{Sch2005} algorithm. This algorithm is a modified local correlation tracking algorithm that also accounts for the contraction, dilation or rotation of the moving magnetic features and computes a velocity field that is consistent with the vertical component of the induction equation.  For this analysis, we used a full width at half maximum of the resolving window of 5 arcsec. 

Figure~\ref{vel} shows the velocity field we derived at a few selected instants in time. We observed persistent photospheric shearing motions around the negative polarity of the northern sunspot. These shearing motions are globally directed toward the PIL of the northern sunspot. However, the northern part of the negative flux distribution is actually subjected to shearing motions directed to the northwest, while the central part of negative flux distribution is subjected to east-west-aligned shearing motions. This trend changes few hours before the filament eruption: the shearing motions are then mostly directed to the southwest (see Fig.~\ref{vel}, bottom panel). 

These measured flows strongly support the hypothesis that the observed flux cancellation is a consequence of the convergence shearing motions towards the PIL of the active region.

Figure~\ref{vel} also shows that the central negative flux distribution is subjected to persistent, southwest-directed shearing motions, while this is not the case for the southernmost negative polarity, which does not display any significant change in its position. As a consequence, by August 3 at 19:12 the two polarities merge and part of the pre-existing (positive) magnetic field is cancelled. These persistent shearing motions and the associated flux cancellation may have been the responsible for the intense flaring activity in the southern part of the active region (see Sect. \ref{Sec:Obs}).

\begin{figure*}
\begin{center}
\includegraphics[width=.9\textwidth]{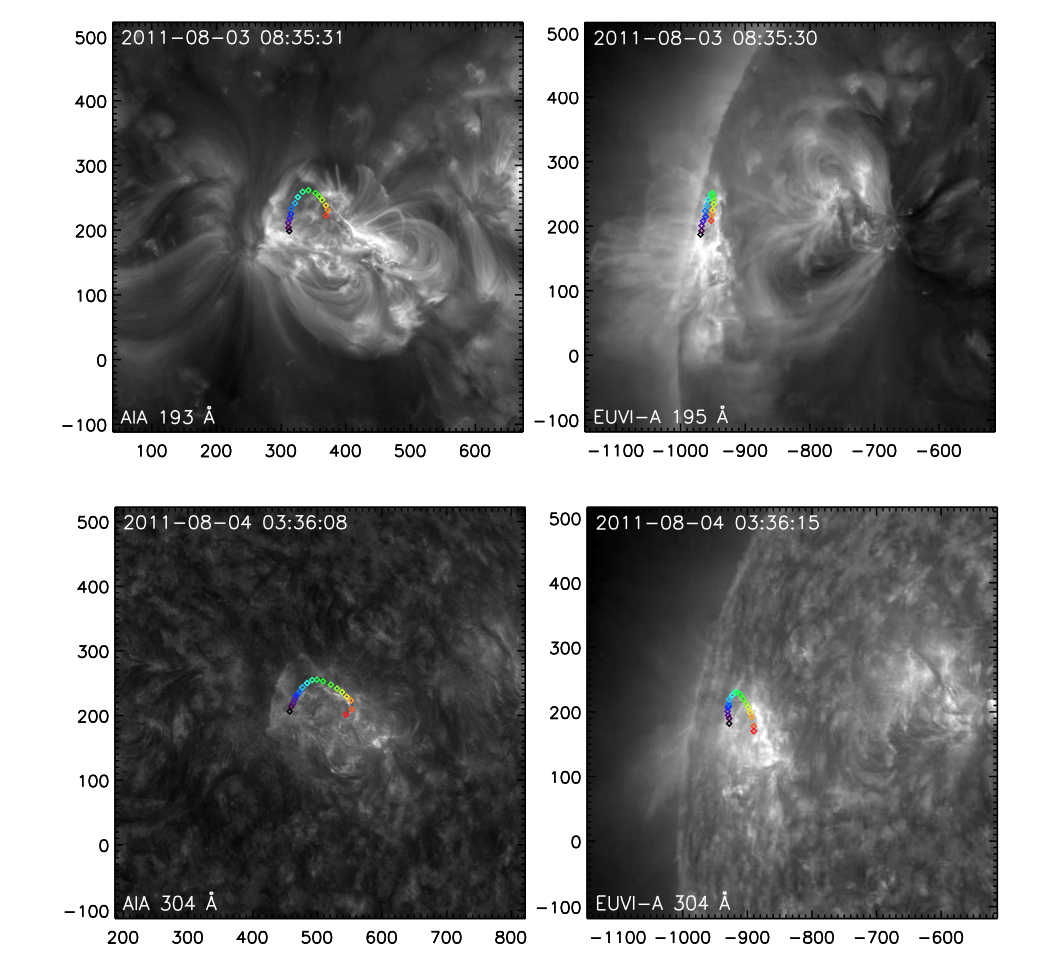}
\caption{Three-dimensional reconstruction of the filament (color diamonds) projected on AIA~193~\AA\,  and EUVI-A~195~\AA\,  images obtained on 2011~August~3 at 08:35~UT (top panels), and on  AIA~304~\AA\, and EUVI-A~304~\AA\,images taken on 2011~August~4 at 03:36~UT (bottom panels). Diamonds of the same color identify the corresponding structures in each image pair. The $x$ and $y$ scales are arcseconds.}
\label{recon}
\end{center}
\end{figure*}

\subsection{3D reconstruction and decay index}


To verify the stability properties of the filament with respect to the torus instability we used the following procedure: First, we reconstructed the three-dimensional position of the filament at several different times. Second, we computed potential magnetic field extrapolations and calculated the decay index of the magnetic field at approximately the same times. Finally, we compared the height of the reconstructed filament with the height at which the decay index of the field exceeds the critical value $n_{crit}$. \cite{Dem2010} have shown that for non-circular, non-fully neutralized current paths the critical value of the decay index $n_{crit}$ is typically in the range 1.2--1.5.

We reconstructed the three-dimensional position of the filament using the triangulation routine \textit{scc\_measure.pro} provided with \textsf{SolarSoft}. This routine, after reading in a pair of images, allows the user to map the line-of-sight of a point selected in one image of the pair into the field of view of the second image \cite[the so-called epipolar line;][]{Inh2006}. After the user identifies the intersection between the projected line-of-sight and the feature of interest, the program can triangulate the feature's three-dimensional location. Because the filament was behind the limb when viewed from STEREO-B, we applied this routine to images from SDO and STEREO-A. 

We performed three reconstructions: first, on August~3 at 08:35, before the M6.0 flare; second, on August~3 at 21:36, about six hours after the M6.0 flare; and, third, on August~4 at 03:36, just before the onset of the filament eruption. For the first reconstruction we used AIA~193~\AA\, and EUVI~195~\AA\, images, because the filament was not clearly visible in EUVI~304~\AA\, images due the bright facula in front of it. However, for the other two reconstructions we used  AIA~304~\AA\, and EUVI~304~\AA\, images, because the contours of the filament were best visible in this wavelength. For each of the image pairs, we identified 15--20 points along the filament for the corresponding reconstruction. 

Figure~\ref{recon} shows the reconstructed filament (cross-referenced colored diamonds) projected back onto the corresponding AIA (left panels) and EUVI (right panels) images. The top panels show the reconstructed position of the filament on August~3 at 08:35 while the bottom panels show the reconstructed position of the filament on August~4 at 03:36. The reconstruction shows that the filament is significantly inclined, that is, the plane containing the axis of the filament is almost parallel to the solar surface. The filament had a maximum height of about 8--10~Mm before the M6.0 flare and of about 14--18~Mm at the moment of the eruption.

\begin{figure*}
\begin{center}
\includegraphics[width=.95\textwidth]{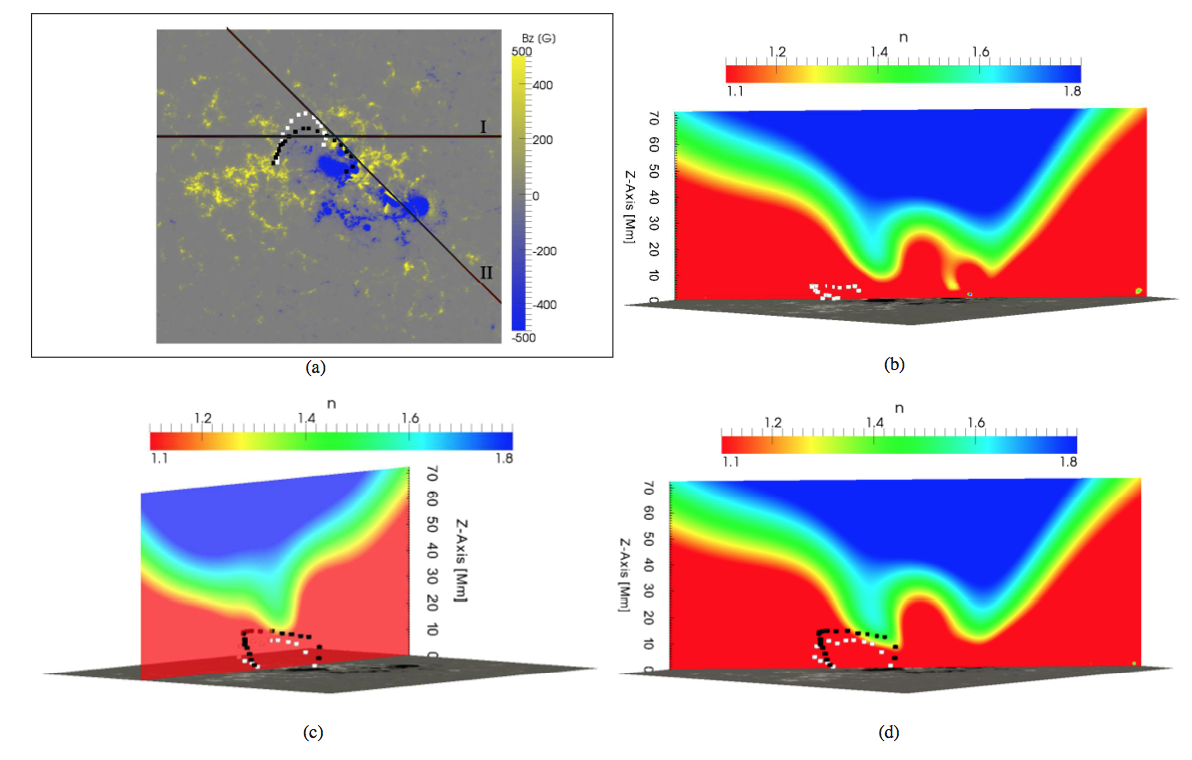}
\caption{ (a) HMI magnetogram of the active region taken on 2011~August~4 at 03:36~UT. The two lines indicate the projections on the magnetogram of the two planes used to display the decay index. The white cubes indicate the reconstructed position of the filament on August~3 at 08:35~UT, while the black cubes highlight the reconstructed position of the filament on August~4 at 03:36~UT. (b) Plot of the decay index for the potential field extrapolation relative to the HMI magnetogram taken on August~3 at 06:00~UT. The decay index is plotted along the plane passing through line II of Figure (a), that is, along part of the axis of the reconstructed filament (white cubes). (c)-(d) Plot of the decay index for the potential field extrapolation relative to the HMI magnetogram from August~4 at 03:36~UT and along the planes passing through lines I and II of Figure (a).} The white and black cubes highlight the three-dimensional position of the filament on August~3 at 21:36~UT and on August~4 at 03:36~UT, respectively. 
\label{decay}
\end{center}
\end{figure*}

\begin{figure}
\begin{center}
\includegraphics[width=.45\textwidth]{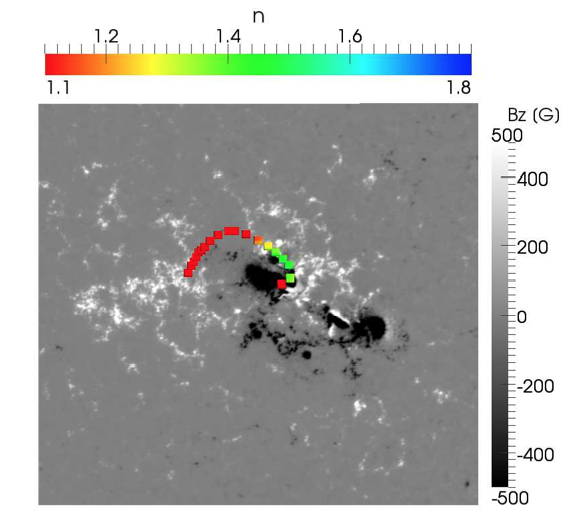}
\caption{HMI magnetograms of the active region taken on 2011~August~4 at 03:36~UT together with the reconstructed position of the filament on August~4 at 03:36~UT. The reconstructed filament is color-coded with the decay index computed from the potential field extrapolation. North is at the top and west to the right.}
\label{decay-color}
\end{center}
\end{figure}

The decay index for a current ring of major radius $R$ and embedded in an external field $B_{ex}$, is defined as:
\begin{equation}
n=-R \frac{d}{dR} \left( \ln B_{ex} \right),
\end{equation}
where $R$ is the flux rope major radius and  $B_{ex}$ is the magnetic field external to the flux rope (see Introduction).

With some approximations, Eq.~(1) can be used to calculate the decay index based on solar observations. First, we estimate the major radius of the flux rope using its peak height above the photosphere \citep{Tor2007,Fan2007,Aul2010} and, second, we use the potential magnetic field as a proxy for the external magnetic field \citep{GuoY2010}. To derive the potential magnetic field for the given HMI magnetogram we use the extrapolation method of \cite{Ali1981}. This method, given the normal component of the magnetic field at the photosphere --- that is, at $z=0$ --- uses the Fast Fourier Transform~(FFT) to calculate the solution of the potential field equation, $\nabla \times \mathbf{B} =0$, in the semi-infinite space $z > 0$. 

After computing the decay index for the extrapolated potential magnetic field, we insert the reconstructed filament into the extrapolation domain. This approach allows us to compare the three-dimensional position of the filament to the profile of the decay index. 

Figure~\ref{decay} shows the computed decay index along two selected planes for the magnetic field extrapolations performed using the magnetograms obtained on August~3 at 06:00 and on August~4 at 03:48. The yellow-green region corresponds to the location where the decay index is $n \simeq 1.3 -1.5$: the height where the system becomes nominally torus unstable. 

Figure~\ref{decay}B shows the decay index along a plane passing through the compact PIL of the active region (line II of Fig.~\ref{decay}A) for the extrapolation on August~3 at 06:00.  It is clear from the figure that the region close to the compact PIL is the most unstable one: in this region the decay index reaches the critical value at a smaller height. This behaviour is also confirmed by Figs.~\ref{decay}C and \ref{decay}D,which show the decay index along the planes passing through lines I and II of Fig.~\ref{decay}A for the extrapolation on August~4 at 03:48. The behavior of the decay index is a consequence of the topology of the active region. In fact, at least close to the filament's location, the magnetic field of the active region can be approximated by two dipoles: first, a large dipole composed of the dispersed positive magnetic field of the facula and the negative polarity of the northern sunspot and, second, a compact dipole composed of the negative polarity of the northern sunspot and the positive polarity to the west of it. The field generated by the compact dipole drops faster with height than the field generated by the large-scale dipole. As a result, the decay index in the proximity of the compact dipole increases more rapidly.

Figure~\ref{decay} also shows that the decay index does not change significantly between the times of first and the second extrapolation. However, after the M6.0 flare, the morphology of the filament changes and the filament extends into the region where the field is more unstable. This change in morphology, and the increased instability of the filament associated with it, is clearly visible when we compare Fig.~\ref{decay}B with Figs.~\ref{decay}C and \ref{decay}D. 

During the M6.0 flare, the filament (white cubes in Fig.~\ref{decay}B) is stable with respect to the torus instability; the reconstructed filament is well below the height where the decay index reaches the critical value $n_{crit} \simeq 1.3-1.5$. However, after the M6.0 flare --- more precisely on August 3 at 21:36 --- the filament (white cubes in Figs.~\ref{decay}C--\ref{decay}D) approaches the region where the decay index of the field is close to the critical value $n_{crit}$. The filament remains in this location until August 4 at 03:36 UT when it erupts. 

The three-dimensional reconstruction of the filament on August~4 at 03:36 (black cubes in Figs.~\ref{decay}C-\ref{decay}D) shows that at the moment of the eruption the filament is in a region where the decay index is in the range of the critical value for the onset of the torus instability.

\section{Discussion}

The filament eruption that occurred on 2011~August~4 is a compelling example of the complexity of coronal dynamics, in which several different phenomena  can all contribute to the initiation of a single CME. In this context it is worthwhile to disentangle which phenomena are essential to causing the eruption -- that is, what, exactly, triggers the eruption --- and which phenomena help facilitate the eruption's onset by bringing the system to a point where the trigger mechanism can work. The numerical MHD simulation performed by \cite{Aul2010} shows that the tether-cutting reconnection driven by flux dispersion can facilitate an eruption by bringing the flux rope to a height where the decay index of the overlying field is larger than the critical value for the onset of the torus instability. But in their simulation, the torus instability is the final trigger mechanism for the eruption.	

In the previous sections, we discussed the morphology of the observed filament and its relation to the photospheric magnetic field. As we show in Fig.~\ref{morph}, before the M6.0 flare, two filaments (white arrows) are visible both in H$_{\alpha}$ and  AIA 193~\AA\, images. We believe that the two filaments are actually part of the same flux rope that extends from the positive magnetic field of the facula to the southern part of the sunspot. This flux rope also has a significant amount of its flux anchored in the northern part of the sunspot. Complex flux ropes with several footpoints have been discussed by \cite{Dem1996} and modeled in numerical MHD simulations \citep{Zuc2012c}. In this case, the additional footpoint of the flux rope may simultaneously support the plasma that constitutes the filaments and introduce significant line-tying effects. 

When part of the smaller filament erupted on August~3 at 13:17, resulting in a M6.0 flare, the larger filament was nonetheless unperturbed. To better understand why this filament was relatively unaffected by a significant flare just to its south, we reconstructed the three-dimensional structure of the larger filament a few hours before the flare and compared its position with the decay index for the potential magnetic field. Our results show that the bigger filament was, at that time, in a region that is stable to the torus instability. This may explain why the larger filament did not erupt despite the occurrence of the M6.0 flare close to its western foot point.

On the other hand, the reconnection associated with the flare changed the morphology of the filament (see Fig.~\ref{morph}D). This reconnection appears to have reduced the line-tying field associated with the flux rope and allowed the plasma to fill the entire flux rope. After the restructuring, the filament grew in length.  While it still spanned its original location between the eastern facula and edge of the northern sunspot, it now extended along nearly the full length of the PIL of the compact dipole composed of the positive-negative polarity of the northern sunspot. This was a fundamental step in the filament's evolution towards eruption. In fact, in the region into which the restructured filament then extended, the overlying field scaled much more rapidly with height and, as a consequence, the decay index approached its critical value at lower heights. Figures~\ref{decay}C--\ref{decay}D show that after the M6.0 flare, on August~3 at 21:36, the filament (white cubes) was very close to the torus instability region. However, no eruption occurred until August~4 at 03.48; the filament remained stable for at least six hours.

The 3D reconstruction of the filament as it was just before the eruption, on August~4 at 03:36, suggests that during the hours that preceded the eruption, the filament underwent a further rise, definitely reaching a height at which the decay index was in the region of the critical value for the onset of the torus instability. During the eight to ten hours preceding the eruption, we observed flux cancellation driven by convergence motions towards the compact PIL of the northern sunspot. Similar to the dynamics in the MHD simulations of \cite{Zuc2012c}, these convergence motions and the associated flux cancellation may have induced tether-cutting reconnection below the filament, which subsequently drove its slow rise. This slow rise abruptly changed at 03:48 when the filament suddenly erupted as a result of torus instability.

Figure~\ref{decay-color} shows a top view of the three-dimensional reconstruction of the filament, color coded using the decay index value at each reconstructed point and projected onto the HMI magnetogram obtained on August 4 at 03:36. At that time, the filament was located directly above the complex PIL that extended from the positive magnetic field of the facula up to the southern part of the negative polarity of the northern sunspot. It is clear in the figure that the part of the filament that had become torus-unstable at that time (green in the color scale of the figure) is the same part that was directly above the compact PIL of the northern sunspot. Furthermore, from Figure~\ref{event}F it is clear that this is also the location from which the eruption first began. We believe this is compelling evidence that the torus instability was the trigger of the 2011~August~4 filament eruption.

\acknowledgements

This research was funded by projects 3E120343 (PDMK/12/138, KU Leuven), 1272714N (FWO Vlaanderen), GOA/2009-009 (KU Leuven), G.0729.11 (FWO Vlaanderen) and C 90347 (ESA Prodex 9) and the Interuniversity Attraction Poles Programme initiated by the Belgian Science Policy Office (IAP P7/08 CHARM). Additional support for D. B. S. and L. A. R. was provided by the Belgian Federal Science Policy Office through the ESA-PRODEX program, grant No. 4000103240 and by Grant FP7/2007-2013 from the European Commissions Seventh Framework Program under the agreement eHeroes (Project No. 284461). The research leading to these results has also received funding from the European Commission's Seventh Framework Programme (FP7/2007-2013) under the grant agreements
SOLSPANET (Project No. 269299, www.solspanet.eu), SPACECAST (Project No. 262468, fp7-spacecast.eu), and SWIFF (Project No. 263340, www.swiff.eu).
We acknowledge the use of data from STEREO/SECCHI and Big Bear Solar Observatory/New Jersey Insitute of Technology. 
The AIA and HMI data are courtesy of NASA/SDO and the AIA and HMI science teams.

\bibliographystyle{aa}
 \bibliography{biblio.bib}

\end{document}